\newcommand{\comment}[1]{}
\comment{

}

\newcommand{\ignore}[1]{}

\documentstyle[11pt,epsf]{article}

\newcommand\be{\begin{equation}}
\newcommand\ee{\end{equation}}
\newcommand\bea{\begin{eqnarray}}
\newcommand\eea{\end{eqnarray}}
\newcommand\half{{\textstyle{1\over2}}}
\newcommand\quarter{{\textstyle{1\over4}}}
\newcommand{\smfrac}[2]{{\textstyle{{#1}\over{#2}}}}

\def\pip{\pi^{\prime}}
\def\sigmap{\sigma^{\prime}}

\def\psibar{{\bar \psi}}

\def\Asl{A \llap \slash}
\def\dsl{\partial \llap \slash}

\def\khat{\hat k}
\def\ktilde{\tilde k}

\begin{document}

\begin{titlepage}
\noindent BUHEP-94-3, BROWN-HET-920\hfill \today \\
\begin{center}

{\Large\bf Chirally Extended Quantum Chromodynamics}

\vspace{1.5cm}
{\bf Richard C. Brower$^{(1)}$, Yue Shen$^{(1)}$ and Chung-I Tan$^{(2)}$}\\
\end{center}
\vspace{1.0cm}
\begin{flushleft}
{}~~$^{(1)}$Department of Physics, Boston University, Boston, MA 02215, USA \\

{}~~$^{(2)}$Department of Physics, Brown University, Providence, RI 02912,
USA\\
\end{flushleft}
\vspace{1.5cm}

\abstract{
We propose an extended Quantum Chromodynamics (XQCD) Lagrangian in which the
fermions are coupled to elementary scalar
fields through a
Yukawa coupling which preserves chiral invariance.  Our principle motivation
is to find a new lattice formulation for QCD which avoids the source of
critical slowing down usually encountered as the bare quark mass is tuned to
the chiral limit.  The phase diagram and the weak coupling limit for XQCD are
studied.  They suggest a conjecture that the continuum limit of XQCD is the
same as the continuum limit of conventional lattice formulation of QCD.  As
examples of such universality, we present the large N solutions of two
prototype models for XQCD, in which the mass of the spurious pion and sigma
resonance go to infinity with the cut-off. Even if the universality conjecture
turns out to be false, we believe that XQCD will still be useful as a low
energy effective action for QCD phenomenology on the lattice.  Numerical
simulations are recommended to further investigate the possible benefits of
XQCD in extracting QCD predictions. }
\vfill
\end{titlepage}

\noindent{\bf I. Introduction}

The lattice cut-off approach to Quantum Chromodynamics has developed into an
increasingly powerful and sophisticated discipline over the last ten years.
The numerical simulations of lattice QCD have improved our understanding of
hadron dynamics and produced interesting nonperturbative results
\cite{Lattice92}: for example, the spectrum of hadrons, the strong interaction
corrections to the weak decay matrix elements and the high temperature behavior
of the gluonic plasma.

However, most results obtained so far are in the quenched approximation where
the internal dynamical fermion loops are neglected.  A major difficulty for
dynamical fermion simulations in the standard approach is the problem
of taking the chiral limit for the quarks by tuning the bare quark mass almost
to zero.  In the numerical simulations that involve dynamical fermions, one
needs to calculate the inverse of the fermion matrix in every sweep over the
lattice.  The fermion matrix has zero eigen-values when the bare quark mass is
exactly zero. Therefore, when the bare quark mass is tuned to the chiral
limit, the number of iterations needed to obtain the inverse of the fermion
matrix with a fixed accuracy becomes exceedingly large and enormous
computer resources are required\cite{TERAFLOPS}.

\bigskip

\underbar{\sc XQCD Lagrangian:} As a possible scheme to speed up the QCD
simulation with
dynamical fermions, we propose a chirally invariant extended lattice cut-off
scheme for QCD (XQCD or eXtended QCD). The Lagrangian, in the continuum
notation, can be written as
\be
{\cal L}_{XQCD}(x) = {\cal L}_{QCD}(x) + {\cal L}_{\Sigma}(x) + y \psibar
( \sigma + i \gamma_5 {\vec \tau} \cdot {\vec \pi}) \psi ~,
\label{eq:xqcd}
\ee
where
\be
{\cal L }_{QCD}(x) = \quarter Tr [F_{\mu\nu} F_{\mu\nu}]
+ \bar \psi (\dsl + ig \Asl) \psi + \bar \psi M \psi  ~,
\label{eq:qcd}
\ee
is the usual QCD Lagrangian and ${\cal L}_{\Sigma}(x)$ is the Lagrangian for
the (linear) sigma model\footnote{Alternatively we may use the non-linear
sigma model in the bare action. As is well known, it is just the $\lambda \to
\infty$ limit of the linear sigma model and has the same continuum limit.}
\be
{\cal L}_{\Sigma} =  \half \partial_\mu \sigma \partial_\mu \sigma +
\half \partial_\mu \vec \pi \partial_\mu \vec \pi + \half m^2 (\sigma^2 +
\vec \pi^2) + \smfrac{1}{24} \lambda (\sigma^2 + \vec \pi^2)^2 ~.
\label{eq:sigma}
\ee
The fermions $\psi^a_\alpha(x)$ are Dirac spinors with color index $\alpha =
1, 2, ..., N_c$ and flavor index $a= 1, 2$.  The Dirac matrices $\gamma_\mu$
are hermitian in Euclidean space with commutators $\{\gamma_\mu, \gamma_\nu\}
= 2\delta_{\mu\nu}$.  For simplicity we have written down the action for the
two flavor case.  The scalar field $(\sigma, {\vec \pi})$ is color singlet and
couples to the fermions in a chirally invariant way: simultaneously rotating
both the scalar field and the left component $\psi_L = \half (1 -
\gamma_5)\psi$ and right component $\psi_R = \half (1 + \gamma_5)\psi$, the
action possesses a $SU_L(2) \times SU_R(2)$ chiral symmetry in the zero mass
limit, $M \rightarrow 0$. Because we expect that the axial $U(1)$ symmetry of
${\cal L}_{QCD}$ will be broken by the triangle anomaly, we have not included
the
axial $U(1)$ symmetry in the scalar sector.  At low energies
the symmetry is further broken down to vector $U(2)$.

Why might we expect XQCD to work better for simulations with dynamical
fermions?  As we will show in Section III, the QCD chiral condensate will
induce a nonzero vacuum expectation value (VEV) for the scalar field.
Therefore, the continuum limit is approached with a finite VEV for the scalar
field. The VEV, in turn, will provide a nonzero mass to the quarks through the
Yukawa coupling term: a mass resembling in some way the constituent mass in the
quark model. We expect that this nonzero constituent quark mass will make the
inverse fermion matrix calculation convergent. Indeed, if we set the gauge
coupling to zero ($g=0$), the model in Eq. (\ref{eq:xqcd}) becomes the
Higgs-Yukawa
model which has been studied extensively in the past few years \cite{fermion}.
The simulations of the Higgs-Yukawa model with dynamical fermions were found
to be well behaved in the broken symmetry phase. We do not expect that
including the gauge field will change qualitatively the behavior of the
fermion determinant.

Assuming that the numerical simulations of XQCD with dynamical fermions is
much easier than the conventional Wilson lattice QCD action, we need to know
in what context XQCD can be used as a substitute of the Wilson QCD action.
Obviously, by coupling the scalar field to the QCD action, we have introduced
extra degrees of freedom at the cut-off scale.  It is less clear, however,
whether these extra degrees of freedom will remain in the continuum limit when
the cut-off is removed: $\Lambda \to \infty$.  Based on the argument given in
this paper, we make the following conjecture:

\begin{itemize}
\item The extra degrees of freedom will remain at the cut-off scale
and be frozen out in the continuum limit. The physical spectrum,
the low and high energy behavior of XQCD will be the same as QCD.
In terms of the language of the renormalization group (RG), XQCD is
in the same universality class as the Wilson QCD action.
\end{itemize}

In our first attempt to search for the continuum limit, we study the phase
diagram and the weak coupling limit for XQCD. We show that there is only one
critical surface in XQCD.  At least in the weak coupling region, there is a
single fixed point on the critical surface. If we insist on removing the
cut-off completely, the renormalized $y$ and $\lambda$ will vanish and the
theory is reduced to the pure gauge theory in the weak coupling limit.

To study possible double counting problem due to our introduction of extra
degrees of freedom in the cut-off theory, we study two toy models that share
some common feature with XQCD. We find that these extra degrees of freedom
have masses at the cut-off scale and therefore will be removed in the
continuum limit.  These results are suggestive that the above conjecture may
hold.  However, the method used to solve the toy models can not be applied to
XQCD.  The conjecture needs to be tested in numerical simulations.

If the extra degrees of freedom do remain in the spectrum after taking the
continuum limit and the above conjecture turns out to be false, the XQCD model
can still be used for QCD phenomenology at energy scale lower than the
spurious particle mass scale. In this case, XQCD is similar in some ways to
the Georgi-Manohar (GM) or chiral quark model which has been proposed as a low
energy phenomenological model for QCD\cite{GM}. However, the GM model has the
additional assumption that the chiral symmetry breaking scale $\Lambda_{\chi
sb}$ is much higher than the confinement scale $\Lambda_{QCD}$, so that there
are intermediate energies at which pions and unconfined quarks coexist:
$\Lambda_{QCD} < < \Lambda_{\chi sb}$. Unfortunately no such energy interval
appears to exist. For example, numerical simulations of QCD at finite
temperature \cite{HTQCD} have demonstrated that deconfinement and chiral
symmetry restoration happen at the same temperature or very nearly the same
temperature: $\Lambda_{QCD} \sim \Lambda_{\chi sb}$. In XQCD confinement and
chiral breaking scales need not be separated, but in this case XQCD is
not more tractable analytically than QCD itself.  Instead we
expect XQCD's advantage lies in more efficient numerical simulations.

\ignore{
As an additional motivation for XQCD, we note that precisely for the chiral
properties, where effective chiral Lagrangians give excellent results, lattice
QCD has its greatest difficulties.  Therefore even on purely phenomenological
grounds, it is natural to ask if there is a hybrid lattice formulation which
begins with a better representation of light quark physics.
}

The organization of the paper is as following: In Section II, we give a
heuristic argument that the XQCD action is a natural extension of the Wilson
lattice QCD action. In section III, we study the phase diagram and show that
there is a single critical surface where one may define the continuum limit.
We show in Section IV in the weak coupling limit that there is only one fixed
point on the critical surface. In Section V, we study the double counting
problem.  We solve two toy models as prototypical versions of XQCD exactly in
the large $N$ limit and show that the extra degrees of freedom are frozen out
in the continuum limit.  Finally, we discuss the fermion doublers in Section
VI.

\newpage
\noindent{\bf II. Extended Lattice QCD}

In general the action for a lattice regulated field theory is not unique.
Any member of a large (universality) class of cut-off Lagrangians,
which share the same basic symmetries of the underlying continuum theory,
will give rise to the same continuum theory. One particular choice of the
lattice action has been written down by
Wilson \cite{Wilson}
\be
{\cal S }_{QCD} = {2 N_c \over g_0^2} \sum_{x, \mu < \nu} (1 - {1\over N_c}
Re \; Tr U_{\mu\nu}(x) ) + \half \sum_{x,\mu} \psibar(x) \gamma_\mu \left[
U_\mu(x)\psi(x+\mu)
-U_\mu^\dagger (x-\mu)\psi(x-\mu)\right]~,
\label{eq:wilson}
\ee
On the lattice, the gauge variable $U_\mu(x) = \exp (ig A_\mu(x))$ are defined
as group elements on each link between $x$ and $x+\mu$ and the gauge action is
formed by the ordered product of links $U_{\mu\nu}(x) = U_\mu(x) U_\nu(x+\mu)
U_\mu^\dagger(x+\nu) U_\nu^\dagger(x)$ around the elementary plaquette in the
$\mu-\nu$ plane.  At present we have not included either the bare quark mass
term or the
Wilson term that is used to remove the doublers in the continuum limit, both
of which explicitly breaks the chiral symmetry. They complicate the
discussion, without affecting the validity of our conclusions.  We defer the
discussion about doublers in Section VI.

The Wilson QCD action has a UV fixed point at $g_0 = 0$. By tuning the bare
gauge coupling to the UV fixed point $g_0 \to 0$, the lattice correlation
length, $\xi$, will increase and eventually we get into the scaling region
where the lattice theory describes a continuum QCD theory with renormalized
coupling $g$.  In the continuum limit, besides the dimension four renormalized
operators that are explicitly written out in the continuum QCD Lagrangian, we
also get all the higher dimensional operators, that respect the gauge and
chiral symmetries, with coefficients determined as functions of $g$. For
example, for two flavors we can have the dimension six four fermion operator
\be
O_6(x) = \left(\psibar\psi\right)^2 - \left(\psibar\gamma_5{\vec \tau}
\psi\right)^2~,
\ee
with a coefficient $G_R = G_R(g) + O(\xi^{-2}\ln^q\xi)$ ($q$ is an integer),
that is completely
determined by the renormalized coupling, $g$. This
operator respects $SU(2)$ chiral invariance but breaks $U(1)$
invariance\cite{tHooft} as a consequence of instanton effects (On the
lattice, $U(1)$ invariance will also be explicitly broken by the Wilson term,
which is used to remove doublers and to reinstate the chiral anamoly).  One
can also add this same four fermion operator to the bare QCD action with an
arbitrary coefficient $G_0$,
\be
{\cal L}_{QCD}^{Lat} \to {\cal L}_{QCD}^{Lat} + G_0 O_6(x)~.
\label{eq:wilsonnjl}
\ee
According to the RG theory \cite{WK,Polchin}, modulo $O(\xi^{-2}\ln^q\xi)$
terms, we can always match the renormalized theory of the actions in Eq.
(\ref{eq:wilson}) and Eq. (\ref{eq:wilsonnjl}) such that the effect of the
four fermion operator in Eq. (\ref{eq:wilsonnjl}) can be absorbed in a
suitable choice of the bare gauge coupling in Eq. (\ref{eq:wilson}). In this
sense the higher dimensional operators are ``irrelevant'': the actions in Eq.
(\ref{eq:wilson}) and Eq. (\ref{eq:wilsonnjl}) belong to the same universality
class and their continuum limits are equivalent.

Alternatively, we may replace the Lagrangian in Eq. (\ref{eq:wilsonnjl}) by
the Lagrangian
\be
{\cal L}_{QCD}^{Lat} + \half (\sigma^2(x) + {\vec \pi}^2(x) )
+ y_0 \sum_x \psibar(x)\left (\sigma(x) + i\gamma_5 {\vec \tau}\cdot
{\vec \pi}(x)\right) \psi(x)~,
\label{eq:ExQCD}
\ee
with $y_0 = \sqrt{2G_0}$. So far we are merely exploiting a mathematical
identity, since integration over the Gaussian fields $(\sigma,{\vec \pi})$
for the new action gives back the same partition function as the old action.
But now we may imagine doing a RG transformation on the new action
by integrating out all the field component, $\psi, \bar \psi,
A_\mu$ and $(\sigma,{\vec \pi})$, within a momentum layer $ e^{-t}\Lambda < p <
\Lambda$. This may be done on the lattice by selectively thinning out the
lattice variables. After this integration, all operators that are absent in
Eq. (\ref{eq:ExQCD}) but allowed under the chiral and gauge symmetries will be
generated.  In particular, if we write down explicitly only the ``relevant''
operators, we get
\begin{eqnarray}
{\cal L}_{QCD}^{Lat} &-& \kappa \sum_{\mu} \sigma_a(x) \left[\sigma_a(x+\mu)
+ \sigma_a(x-\mu)\right] +  \sigma^2_a(x) + \lambda_0 (\sigma_a^2(x)
-1 )^2 \\ \nonumber
&+& y_0 \; \psibar(x)\left (\sigma(x) + i\gamma_5 {\vec \tau}\cdot
{\vec \pi}(x)\right) \psi(x)~,
\label{eq:latxq}
\end{eqnarray}
where for conciseness we have used tensor notation for the chiral four vector:
$\sigma_a = (\sigma,{\vec \pi})$.  After a proper rescaling of the scalar
kinetic term, this action becomes precisely a lattice regularized version of
the XQCD action in Eq. (\ref{eq:xqcd}).

At $g_0 = 0$, Eq. (\ref{eq:ExQCD}) becomes the Nambu-Jona-Lasinio (NJL) model.
In the large $N$ limit \cite{BHL,UCSD}, the dynamical generation of the
kinetic term and the $\lambda (\sigma_a^2)^2$ term has been explicitly
demonstrated for the NJL model.  Furthermore, allowing a large class of higher
dimensional irrelevant operators in the bare Lagrangian, it has been
demonstrated that there is a one-to-one matching relation between the
generalized NJL model and the Higgs-Yukawa model
\cite{UCSD} ($g_0=0$ limit of XQCD in Eq. (\ref{eq:xqcd}) or Eq. (8)).
For XQCD, where
$g_0 > 0$, the large $N$ limit is no longer simple and it is difficult to find
out if matching conditions between Eq. (8) and the Wilson QCD action with
higher dimensional irrelevant operators exist.  However, because the
conclusions of Ref \cite{UCSD} is based more on the general principles of the
RG \cite{UCSD,Zinn} rather than on the technicalities of the large $N$ limit,
we expect that our argument leading to Eq. (8) is  qualitatively correct.

Of course, without knowing the matching condition, we do not know if this
correspondence between XQCD and the Wilson lattice QCD with irrelevant
operators will hold for {\it arbitrary} values of $\beta, \kappa, y_0$ and
$\lambda_0$
in Eq. (8). In order to search for all possible continuum theory
defined by the lattice XQCD action, we need to study its phase diagram and
the fixed point structure.

\newpage
\noindent{\bf III. Phase Diagram}

Let us take the XQCD model on the lattice as given in Eq. (8) and study its
phase diagram in the $(\beta, \kappa, y_0)$ parameter space.  The value of
$\lambda_0$ is not crucial. Here we take $\lambda_0$ to be a fixed value.
Alternatively, one may take the $\lambda_0 \to \infty$ limit.

For $y_0 = 0$ the scalar field sector is decoupled from the QCD sector and the
phase diagram in this limit is well known.  The scalar sector becomes the
$O(4)$ symmetric model with a critical point at $\kappa_{cr}(\lambda_0)$. For
$\kappa > \kappa_{cr}(\lambda_0)$ the $O(4)$ symmetry is spontaneously broken.
For the QCD sector, the chiral symmetry is broken for all values of $\beta$.
Therefore, as indicated in Figure \ref{fig:phasespace} by the horizontal
dotted line, we have two phases in the $y_0=0$ plane,
\be
<\phi> = 0, \ \ <\psibar\psi> \ne 0, \ \ \kappa < \kappa_{cr}(\lambda_0)~; \ \
\\
<\phi> \ne 0, \ \ <\psibar\psi> \ne 0, \ \ \kappa > \kappa_{cr}(\lambda_0)~.
\ee
In the $g_0 = 0$ ($\beta = \infty$) plane, Eq. (8) becomes the $SU(2) \times
SU(2)$ symmetric Higgs-Yukawa model. The phase diagram of this model and
similar models has been studied extensively both for finite $\lambda_0$ and
$\lambda_0 \to \infty$ limit \cite{HLN,Bock,HJS,fermion}.  We sketch the
result as shown in Figure \ref{fig:phasespace}. In the weak $y_0$ region the
symmetric and broken phase is separated by a second order phase transition
line. This entire critical line is inside the attractive domain of the
Gaussian infrared fixed point. Approaching from the broken symmetry phase, the
continuum limit is a coupled scalar-fermion theory with couplings going to
zero logarithmically with an increasing correlation length.  In the strong
$y_0$ region, there appears another symmetric region separated from the broken
symmetry region by a second order phase transition line.  Around this critical
line in the strong $y_0$ region, the fermions are removed from the spectrum
and the continuum limit is a (trivial) scalar theory \cite{fermion,Petcher}.
For our purpose, we are only concerned with the critical line in the weak
$y_0$ region.

\begin{figure}[ht]
$$
\epsfxsize=10.5cm
\epsfysize=10.5cm
\epsfbox{fig11.ps}
$$
\caption{Phase diagram  for XQCD in the $\beta-\kappa-y_0$ volume. Chiral
symmetry is spontaneously broken except for the shaded  regions in
the $\beta = 1/g^2_0= \infty$ plane. The dotted line
in the $y_0 = 0$ plane is the chiral phase boundary for the
scalar sector when $y_0 = 0$.
\label{fig:phasespace}}
\end{figure}

There is no numerical result available in the region where $(\beta, \kappa,
y_0)$ are all finite.  However an expansion in powers of $y_0$ can be
performed for small values $y_0$.  If we use $<\sigma>$ and $<\psibar\psi>$
as order parameters for chiral symmetry breaking, to first nontrivial order in
$y_0$, we find
\begin{equation}
\langle \sigma \rangle = \langle \sigma \rangle_\Sigma - y_0 \;
\chi_\sigma \; \langle \psibar\psi\rangle_{QCD} + O(y_0^2)~,
\label{eq:vev}
\end{equation}
and
\begin{equation}
\langle\psibar\psi\rangle = \langle\psibar\psi\rangle_{QCD} -
y_0 \; \langle\sigma\rangle_\Sigma \; \chi_{\psibar\psi} + O(y_0^2)~,
\label{eq:psibarpsi}
\end{equation}
where the notation,
\begin{equation}
\langle O \rangle_\Sigma = {\int [d\sigma d{\vec \pi}]  \; O \; e^{-S_\Sigma}
\over
\int [d\sigma d{\vec \pi}] e^{-S_\Sigma} }~, \;\;\;\; \langle O \rangle_{QCD}
= {\int [dU d\psibar d\psi] \; O \; e^{-S_{QCD}}  \over
\int [dU d\psibar d\psi] e^{-S_{QCD}} }~,
\end{equation}
express independent averages in the scalar and QCD sectors of XQCD.
The susceptibilities are defined as follows
\be
\chi_\sigma = \sum_x \left[\langle \sigma(x)\sigma(0)\rangle_\Sigma
- \langle \sigma \rangle_\Sigma^2 \right],
\ \ \ \chi_{\psibar\psi} = \sum_x \left[
\langle (\psibar\psi)_x (\psibar\psi)_0\rangle_{QCD}
- \langle\psibar\psi\rangle_{QCD}^2\right].
\ee
Let us align the direction of the QCD chiral symmetry breaking
along
\begin{equation}
\langle \psibar\psi \rangle_{QCD} \ne 0~,~~\langle \psibar \gamma_5
\psi\rangle_{QCD}
= 0~, \ \ 0 < \beta < \infty~.
\end{equation}
It is then clear from Eqs. (\ref{eq:vev}) and (\ref{eq:psibarpsi}) that XQCD
is in the broken phase $<\sigma> \ne 0, \ <\psibar\psi> \ne 0$ at any $\beta$
and $\kappa$ values in the small $y_0$ region. Intuitively, this result is
easy to understand. A paramagnetic material will have nonzero magnetization in
an external magnetic field. For XQCD, the QCD sector acts as an external
symmetry breaking source to the scalar sector. A similar phenomenon will be
also
seen in toy models in Section V (see for example Figure \ref{fig:toyphase}).

To approach the continuum limit we need to tune the bare parameters such that
\begin{equation}
\langle \sigma \rangle \to 0~, ~~\ \  \langle \psibar\psi\rangle \to 0~.
\end{equation}
Therefore, for small $y_0$ the only critical surface in the $(\beta,
\kappa, y_0)$ parameter space is in the shaded region on $\beta=\infty$ plane
as shown in Figure \ref{fig:phasespace}. We have also explored the strong
coupling plane $\beta = 0$ and the strongly disordered plane $\kappa =
0$ to check that there is no sign of chiral symmetry restoration in those
limits. Although we are investigating the phase plane further
we are in fact quite confident that the only critical
surface lies in the weak coupling plane at $\beta^{-1} = g^2_0 = 0$ as
depicted in Figure \ref{fig:phasespace}.

\newpage
\noindent{\bf IV. Weak Coupling Expansion}

We have identified the critical surface in the previous Section.  A particular
way to reach the continuum limit of XQCD is by holding $\kappa$ and $y_0$
fixed within the bounds of the shaded region in Figure \ref{fig:phasespace}
and tune $\beta \to \infty$. Universality of the critical phenomenon means
that any path that ends at the same point on the critical surface will define
an equivalent continuum theory.  Furthermore, if the whole critical surface
falls in the domain of a single fixed point, the continuum limit should be in
general the same no matter which point on the critical surface is approached.

Perturbation theory allows us to study the fixed point structure in the
small bare coupling parameter space.
Let us define $t = \ln \xi$, with $\xi$ an
appropriate correlation length on the lattice. To approach the continuum
limit, $\xi$ must diverge as the bare action is tuned to the critical surface.
We have derived the  $\beta$ function to the one-loop order for XQCD:
\bea
-{d g^2\over dt} &=& \beta_{g^2} = {g^4\over 8\pi^2}\left({4\over 3}
- {11\over 3}N_c\right)~,
\label{eq:betag}\\
-{d y^2\over dt} &=& \beta_{y^2} = {y^2\over 4\pi^2}\left[(3+2 N_c)y^2
- {3\over 2}{N_c^2-1\over N_c}g^2\right]~,
\label{eq:betay}\\
-{d \lambda\over dt} &=& \beta_{\lambda} = {1\over 4\pi^2}\left[\lambda^2
+4 N_c \lambda y^2 - 24 N_c y^4 \right]~.
\label{eq:betalambda}
\eea
Eqs. (\ref{eq:betag}), (\ref{eq:betay}) and (\ref{eq:betalambda}) reproduce
the essential features of the fixed point structure when the gauge
and the Higgs-Yukawa sectors are decoupled: there is a UV fixed point
for the gauge sector at $g = 0$ and an infrared fixed point for the
Higgs-Yukawa
sector at $\lambda=0, y = 0$ \cite{fermion}.

\begin{figure}[ht]
$$
\epsfxsize=10.5cm
\epsfysize=10.5cm
\epsfbox{fig22.ps}
$$
\caption{Weak coupling renormalization group flow in
the $g^2-\lambda-y^2$ parameter space.\label{fig:weakphase}}
\end{figure}

For XQCD, because $\beta_{g^2}$ and $\beta_{y^2}$ do not depend
on $\lambda$ to leading
order , $g^2(t)$ and $y^2(t)$ can be solved explicitly.  Defining $r
= y^2/g^2$, we get
\bea
g^2(t) &=& {g_0^2 \over 1 - {29 \over 24\pi^2}g_0^2 t}~,\\
r(t) &=& {5\over 54} {1 \over {r_0 + 5/54 \over r_0}\left({g^2\over
g_0^2}\right)^{5/29} -1 }~,
\eea
where $g_0, r_0 = y_0^2/g_0^2$ are the initial values at $t=0$.  Using these
solutions and Eq. (\ref{eq:betalambda}) the RG flow can be easily plotted as
shown in Figure \ref{fig:weakphase}.  The only fixed point is at $g_0 = 0, y_0
= 0, \lambda_0 = 0$. This fixed point is UV attractive in $g^2$ direction and
repulsive in $y^2$ and $\lambda$ direction. The gauge field retains the
asymptotic freedom, which is essential to have a confining theory.
To define a continuum limit, one
sends $t \to \infty$ while adjusting the bare parameters $g_0^2, y_0^2$ and
$\lambda_0$ such that $g^2, y^2, \lambda$ are fixed to a set of prescribed
(physical) value.
However, a careful inspection shows that unless $y^2, \lambda$ are
fixed to zero, there is always a maximum correlation length at which $y_0^2$
or $\lambda_0$ becomes infinite.  This is, of course, the well known
triviality property.\footnote{Curiously, if we had more copies of fermions in
the model, there would be an infrared fixed line along which $y^2$ and
$\lambda$ are nonzero and proportional to $g^2$.  Such infrared behavior has
been explored in the Standard Model \cite{PR}.} Therefore, if we insist on
removing the cut-off the Higgs-Yukawa sector will be decoupled and we get back
to the pure gauge theory.

Although perturbation theory give us the fixed point structure in the
weak coupling region, it does not answer an important question:
by coupling a scalar sector to QCD, have we introduced a new mass scale
parameter independent of $\Lambda_{QCD}$?
This question can not be answered within the framework of perturbation theory.
Because the anomalous dimensions are small around the UV fixed point,
mass parameters introduced in the action are expected to remain independent.
The new mass parameter introduced in XQCD is the mass of scalar fields,
which is controlled by tuning $\kappa$ to $\kappa_c$.

Intuitively we expect the continuum limit of XQCD to be the same as the Wilson
lattice QCD when the critical surface is approached in the region where both
$\kappa$ and $y_0$ are small. In this region the scalar mass is order one in
cut-off units. Integrating out the scalar field will only generate short
ranged nonlocal higher dimensional operators that fall off exponentially at
long
distances and should not change the universality class of the continuum limit.
However, when the critical surface is approached near the critical line of the
Higgs-Yukawa sector in the $\beta=\infty$ plane, the scalars could generate
nonlocal higher dimensional operators with a long range. While the naive
expectation is that this will alter the universality class of the continuum
limit, really one should integrate out the unphysical degrees of freedom,
which may not be equivalent to integrating out the scalar field and may not
generate new relevant nonlocal operators, as we will illustrate with the toy
models in the next Section.

\newpage
\noindent{\bf V. Toy Models}

Compared to QCD, our model in Eq. (\ref{eq:xqcd}) appears to have extra
degrees of freedom. Let us call the physical pion and the sigma resonance in
QCD $\pi$ and $\sigma$ respectively, and the extra degrees of freedom in XQCD
$\pi^\prime$ and $\sigma^\prime$.  Because of the Yukawa coupling in XQCD, the
interpolating fields for the pionic states ($\pi$, $\pi^\prime$) will
be linear combinations of $\psibar(x)\gamma_5 {\vec \tau}\psi(x)$ and
$\pi(x)$; and for the sigma resonances ($\sigma$, $\sigma^\prime$), linear
combinations of $\psibar(x)\psi(x)$ and $\sigma(x)$.  As we have mentioned in
the introduction, when the cut-off is removed, there are two possible
scenarios: (1) $\pi^\prime, \sigma^\prime$ masses remain at the cut-off scale
in the continuum limit. In this case, the extra degrees of freedom are
removed, and the physical spectrum is the same as the QCD spectrum; (2) Both
$\pi^\prime$ and $\sigma^\prime$ are light with the mass scale either set by
$\Lambda_{QCD}$ or set by an independent mass scale.  The extra degrees of
freedom cannot be removed and the continuum limit will not be equivalent to
QCD.

In the following we will study two toy models that share some similarity with
XQCD. We will show that the extra degrees of freedom introduced in the
bare action for these models do not survive the continuum limit.

\underbar{\sc Extended Nambu-Jona-Lasinio Model}:
Our first toy model for XQCD is constructed by replacing the gluonic force by a
chirally invariant four-Fermi interaction of the Nambu-Jona-Lasinio model.
This toy model, which we call the extended Nambu-Jona-Lasinio (XNJL)  model,
has Lagrangian,
\be
{\cal L}_{XNJL} =   {\cal L}_{NJL}  + {\cal L}_{\Sigma}
+ y_0 \psibar ( \sigma + i \gamma_5{\vec \tau} \cdot {\vec \pi}) \psi ~,
\ee
where
\be
{\cal L}_{NJL} = \psibar \dsl \psi + \half {g_0^2 \over \Lambda^2}
[( \psibar \psi )^2
- (\psibar \gamma_5 \vec \tau \psi)^2 ],
\ee
and
\be
{\cal L}_{\Sigma} =  \half \partial_\mu \sigma \partial_\mu \sigma +
\half \partial_\mu \vec \pi \partial_\mu \vec \pi + \half m_0^2 (\sigma^2 +
\vec \pi^2) + \smfrac{1}{24} \lambda_0 (\sigma^2 + \vec \pi^2)^2 ~.
\ee
The  cut-off scale $\Lambda$ is written out explicitly to make all
the couplings dimensionless. For the purpose of our analysis, we work
with a momentum cut-off scheme.  Similar to XQCD, the physical
pion and sigma resonance in this model are linear combinations of
$\psibar\gamma_5 \vec \tau \psi \sim \vec \pi$ and $\psibar\psi \sim \sigma$,
respectively.

\begin{figure}[ht]
$$
\epsfxsize=10.5cm
\epsfysize=10.5cm
\epsfbox{fig33.ps}
$$
\caption{Phase diagram  for XQCD toy model in the $\beta-m^2_0$ plane
for a fix value of $y_0 > 0$.
The chiral symmetry is spontaneously broken except for
the shaded region bounded by the critical line $m^2_0 (g^2_{cr} - g^2_0) =
y^2_0 \Lambda^2$. The dotted lines are the chiral phase boundaries (horizontal
and vertical lines for the $\langle \sigma \rangle$ and $\langle\psibar\psi
\rangle$
order parameters respectively) when $y_0 = 0$ and the two sectors are
decoupled.\label{fig:toyphase}}
\end{figure}

The four fermion term may be replaced by a Gaussian integration over
auxiliary scalar fields, $\phi_o, {\vec \phi}$, in the new Lagrangian,
\be
{\tilde {\cal L}}_{XNJL} = \psibar \dsl \psi
+ \half \Lambda^2  \phi_a^2(x) + \psibar(x)\left[ g_0 \phi_0
+ y_0 \sigma + i \gamma_5 \left(g_0 {\vec \phi} + y_0{\vec \pi}
\right)\cdot{\vec\tau} \right] \psi(x) + {\cal L}_{\Sigma} ~.
\label{eq:XNJL}\ee
In this form the model is easily solved in the large $N_c$ (color) limit.
Since only the linear combination $g_0 \phi + y_0 \sigma$ appear in the Yukawa
term, we define a shifted field $\varphi = g_0 \phi + y_0 \sigma$.  To leading
order in large $N_c$, the effective potential of the model is
\be
V(\varphi, \sigma)/N_c = \half \Lambda^2 \left( {\varphi_a^2\over g_0^2}
+ {y_0^2 \sigma^2_a\over g_0^2} - {2 y_0 \over g_0^2}\sigma_a \varphi_a\right)
+ \half m_0^2 \sigma^2_a +{1\over 24} \lambda_0  (\sigma_a^2 )^2
- 4 \int {d^4p \over (2\pi)^4} \ln (p^2 + \varphi^2_a) ~
\ee
and the gap equations are found by setting the first order
derivative of $V(\varphi,\sigma)$ to zero. Adopting the convention
$\sigma_a=\varphi_a=0$ for $a\neq 0$ and dropping the subscript
for $a=0$, the VEV's are given by
\be {y_0^2\Lambda^2 \over g_0^2}
\sigma - {y_0 \Lambda^2 \over g_0^2 }\varphi + m_0^2 \sigma +
{\lambda_0\over 6}\sigma^3 =0 ~,
\ee
\be
{\Lambda^2 \over g_0^2} \varphi - \frac{y_0 \Lambda^2}{g_0^2} \sigma
- 8 \int {d^4p\over (2\pi)^4} {\varphi \over p^2 + \varphi^2} =0 ~.
\ee
 To
determine the critical surface, we tune $\sigma \to 0, \varphi \to 0$ while
keeping the ratio $\sigma/\varphi$ fixed, and we get \be
m_0^2 \left( g^2_{cr} - g_0^2 \right) = y_0^2 \Lambda^2~,
\ee
where
\be
\Lambda^2/g_{cr}^2 = 8 \int {d^4p \over (2\pi)^4}{1\over p^2}
= {\Lambda^2 \over 2\pi^2}
\ee
is the critical point for the NJL model.  A cross section of this
critical surface at a fixed $y_0$ is plotted in Figure \ref{fig:toyphase}.
Similar
to XQCD, the symmetry breaking of the NJL sector will affect the
scalar sector such that the broken phase is enlarged.  Here, in contrast to
XQCD, in order to form the fermion anti-fermion bound state, the
coupling must be larger than a nonzero critical value, $g_{cr}^2 \ne
0$. For smaller values of the coupling, the XNJL model has a
symmetric phase with no Goldstone modes.

We next turn to the masses in the large $N_c$ limit. First, it
follows from Eq. (\ref{eq:XNJL}) that the fermion mass is related to
the VEV by $m_f=\varphi$. We can calculate various boson masses by
using the second order derivatives of the effective potential evaluated at the
minimum. We note that the ``kinetic" terms for the scalar fields
take on the form
\be
{\cal L}_{XNJL}= \half I_{\Lambda} \partial_\mu \varphi  \partial_\mu \varphi +
\half \partial_\mu \sigma \partial_\mu \sigma + \cdots ,
\ee
where
\be
I_\Lambda = 4 \int {d^4p\over (2\pi)^4}{1 \over (p^2+\varphi^2)^2}
\sim {1\over 4\pi^2}\ln {\Lambda^2\over \varphi^2}~.
\ee
This introduces a  wave-function renormalization for $\varphi$.

Consider the pseudoscalar sector first. The relevant second order derivatives
of
the effective potential evaluated at the minimum lead  to  a two by
two mass matrix\footnote{For simplicity, we define
the masses by the two-point function at zero external momentum.
This is
different from the physical pole mass.}
\be {y_0\over g_0^2}\Lambda^2 \left( \begin{array}{cc} {\varphi\over
\sigma} & -I_{\Lambda}^{-1/2} \\ -I_{\Lambda}^{-1/2} &   {\sigma\over
\varphi}I_{\Lambda}^{-1} \end{array}\right)~. \ee
Diagonalizing this matrix, we get the eigen-values
\be
m_\pi^2 = 0, \ \ \ m^2_{\pip} = {y_0\over g_0^2} \Lambda^2\left({\varphi\over
\sigma}
+ {\sigma\over \varphi}I_{\Lambda}^{-1} \right) ~.
\ee
The fact that only one pion is massless is not surprising at all. Because the
chiral symmetry of NJL model is locked with the symmetry of the sigma
model, there should be only one set of Goldstone boson when the symmetry
is spontaneously broken. What is interesting is that the mass of the
extra pion, $\pip$, stays at the cut-off scale. Therefore, when the
cut-off is removed, only the massless pion will remain in the spectrum.

Using the second order derivatives along the symmetry breaking direction,
we get the $\sigma$ mass matrix
\be
\left( \begin{array}{cc}
{y_0\over g_0^2}\Lambda^2{\varphi\over \sigma} + {\lambda_0\over 3}\sigma^2
 & -{y_0\over g_0^2}\Lambda^2 I_{\Lambda}^{-1/2}\\
-{y_0\over g_0^2}\Lambda^2 I_{\Lambda}^{-1/2} &  {y\over g^2}\Lambda^2
{\sigma \over \varphi}I_{\Lambda}^{-1} + 4\varphi^2
\end{array}\right) ~.
\ee
After diagonalization, we find the mass eigen-values
\be
m_\sigma^2 = {1\over {\varphi\over \sigma} + {\sigma\over
\varphi}I_{\Lambda}^{-1}} \left( 4
 {\varphi^3 \over \sigma} +{\lambda_0\over 3}{\sigma^3\over \varphi}
I_\Lambda ^{-1}\right)~,
\ee
\be
m^2_{\sigmap} = {y_0\over g_0^2}\Lambda^2\left({\varphi\over \sigma}
+ {\sigma\over \varphi}I_\Lambda ^{-1}\right)
+ {\varphi\sigma \over {\varphi\over \sigma} +
 {\sigma\over \varphi}I_\Lambda ^{-1}}
\left({\lambda_0\over 3}+ 4 I_{\Lambda}^{-1}\right)~.
\ee
One finds that the $\sigma$ mass is proportional to the VEV and will
stay in the physical spectrum when the cut-off is removed.
The mass of $\sigmap$, on the other hand, stays at the cut-off scale and
will drop out of the physical theory.  The fact that $\sigmap$ and
$\pip$ masses are the same to the leading order in $\Lambda^2$
reflects the underlying chiral symmetry. Their mass splitting is due
to the symmetry breaking which should be at the VEV scale.

\underbar{\sc Coupled O(N) Models:}
Next we consider another exactly soluble model.  For two flavors $n_f
= 2$ the sigma model is equivalent to an $O(4)$ model. Therefore it
is natural to consider coupled O(N) model in the large $N$ (``flavor'') limit.
The Lagrangian is given by
\be
{\cal L } = \half \partial_\mu\phi_a \partial_\mu \phi_a
+ \half m_1^2 \phi_a^2 + {\lambda_1\over 24}\left(\phi_a^2\right)^2
+ \half \partial_\mu\sigma_a \partial_\mu \sigma_a
+ \half m_2^2 \sigma_a^2 + {\lambda_2\over 24}\left(\sigma_a^2\right)^2~
- y \Lambda^2 \phi_a\sigma_a.
\label{eq:XON}
\ee
We have inserted $\Lambda^2$ to make the coupling $y$ dimensionless.
Since the details of the analysis are very similar to the XNJL model,
we will just state our results: in the broken phase both $\phi$ and
$\sigma$ field acquire VEV, $<\phi_1> = \phi$, $<\sigma_1> = \sigma$.
For the $2N$ degrees of freedom in the bare Lagrangian,
there are $N-1$ (Of course, $N-1$ cannot be distinguished from $N$ in
the large $N$ limit) massless Goldstone bosons, one $\sigma$ resonance
with mass at the scale of the VEV, $N-1$ massive $\pip$ and a
massive $\sigmap$ with mass
\be
m_{\sigmap}^2 \sim m_{\pip}^2 = y\Lambda^2\left( {\phi \over \sigma}
+ {\sigma \over \phi}\right)~.
\ee
Therefore,
when the cut-off is removed, this model is equivalent to a $O(N)$ sigma model
with $N-1$ Goldstone bosons and a $\sigma$ resonance.

Intuitively, it is easy to understand the above result for
the coupled $O(N)$ model
in Eq. (\ref{eq:XON}) in the lattice regularization. The action becomes
\be
S = -\kappa_1 \sum_{x,\mu} \phi_a(x)\phi_a(x+\mu) - \kappa_2\sum_{x,\mu}
\sigma_a(x) \sigma_a(x+\mu) - \sqrt{\kappa_1\kappa_2}
y \sum_x \phi_a(x)\sigma_a(x) + {\rm local~terms}~.
\label{eq:lattice}
\ee
Assuming the linear size of the lattice is $L$, Eq. (\ref{eq:lattice})
becomes an $O(N)$ model on a $L^4\times 2$ lattice with effective hopping
parameter $\sqrt{\kappa_1\kappa_2} y $ in the fifth dimension. When this model
is tuned to the continuum limit, the correlation length $\xi \sim 1/m_\sigma$
becomes divergent. As soon as $\xi \gg 2$ the effective dimension becomes
four and the model is equivalent to a four dimensional $O(N)$ model.

Although the above analysis for the toy models cannot be directly carried
over, one may imagine similar scenario for XQCD in Eq. (\ref{eq:xqcd}) where
the masses of $\pip, \sigmap$ remain at the cut-off scale and are removed in
the continuum limit.  Then the continuum limit of the XQCD model is very
likely to be equivalent to the Wilson lattice QCD action.  However, we can not
exclude the possibility of a second scenario: $\pi^{\prime}, \sigma^{\prime}$
will stay in the physical spectrum.

Let us imagine that at low energy, the QCD chiral condensate can be
represented by a scalar field $\phi_{ij} \sim <\psibar_i\psi_j>/
\Lambda_{QCD}^2$.
Then the Yukawa coupling term in Eq. (\ref{eq:xqcd}) becomes
\be
y \psibar \left(\sigma + i\gamma_5 {\vec \pi}\cdot {\vec \tau}\right)\psi
\to y \Lambda_{QCD}^2 tr \left(\phi_0\sigma + {\vec \phi}\cdot{\vec \pi}
\right)~,
\label{eq:QCDpot}
\ee
e.g., it is similar to the coupled $O(N)$ model in Eq. (\ref{eq:XON}).
However, here the scale in Eq. (\ref{eq:QCDpot}) is $\Lambda_{QCD}$
instead of the
cut-off $\Lambda$ (remember, we have not taken the $\Lambda \to \infty$
limit yet). Therefore, the $\pip$ and $\sigmap$ masses are of the order
\be
m_{\pip}^2 \sim m_{\sigmap}^2 \sim y \Lambda_{QCD}^2 \left(
{<\sigma>\over <\phi_0>} + {<\phi_0>\over <\sigma>} \right)~,
\ee
which will stay in the physical spectrum when the cut-off is removed.
If this scenario turns out to be true, then XQCD in
Eq. (\ref{eq:xqcd}) can be only used as effective low energy
phenomenology model as discussed in the introduction section.
Presumably it might be possible to choose a particular set of bare parameters
such that the ratio $<\phi_0>/<\sigma>$ is either large or small. Then the
$\pip,\sigmap$ masses can still be much higher than $\Lambda_{QCD}$.

\newpage
\noindent{\bf VI. Removing Doublers}

So far for simplicity we have worked with naive fermions.  For the purpose of
investigating how to accelerate dynamical fermion simulations, this is
adequate. However, for realistic situations, we have to deal with the problem
of fermion doublers.  At first, it appears that a Wilson-Yukawa second order
derivative term would be a better choice for removing doublers because it
preserves the chiral invariance. However, in the $g_0 = 0$ limit, this becomes
a version of the Smit-Swift model, which has been studied extensively in the
last few years. In general it was found that a Wilson-Yukawa term either can
not lift the doublers or removes all fermions from the spectrum
\cite{Petcher}. It is unlikely that including the gauge field will change this
situation. Thus we are left with the option to use the original Wilson term
\begin{equation}
S \to S - {r\over 2} \sum_{x,\mu} \psibar(x) \left[ U_\mu(x)\psi(x+\mu)
+ U_\mu^\dagger(x-\mu)\psi(x-\mu) - 2 \psi(x)\right]~.
\end{equation}
The obvious problem with the Wilson term is that it breaks chiral symmetry.
Therefore, internal loops involving the Wilson term will generate divergent
terms that requires adding lower dimensional operators as counter terms.
In the original lattice QCD action, the only possible counter term is
the bare fermion mass term $m_0 \psibar \psi$. With the inclusion of the
scalar fields, however, there are additional counter terms.

Let us illustrate this point in the $g = 0$ limit where the calculation can be
done exactly in the large $N$ limit.  We can write the action as
\begin{equation}
L = \psibar \dsl \psi - {r\over 2} \psibar \partial^2 \psi + y \psibar (\sigma
+
i\gamma_5{\vec \pi}\cdot {\vec \tau})\psi
+ {1\over 2}\partial_\mu\sigma_a\partial_\mu\sigma_a + {m^2\over 2}\sigma_a^2
+ \smfrac{1}{24}\lambda  \left(\sigma_a^2\right)^2~,
\end{equation}
where the kinetic part for the fermion, scalar fields and the Wilson term
are written only symbolically in continuum notation ($\sigma_a =
(\sigma,{\vec \pi})$ as before). Their form on the lattice
are obvious. In the large $N$ limit, it is easy to find the fermion spectrum
\begin{equation}
m_f = yv + 2n{r \over a}, ~\ \ \ n = 0, 1, 2, 3, 4~
\end{equation}
where we have restored the lattice spacing $a$. It is clear that in the
continuum limit $a \to 0$, all the doublers will be removed as long as
$r \sim O(1)$. We are left with a light fermion with mass $m_f = yv$.

Since the Wilson term breaks the chiral invariance, the Ward identities
associated with chiral symmetry are no longer satisfied. One of these chiral
Ward identities is the vanishing of the pion two point function at zero
external momentum $G_{\pi\pi}^{-1}(0) = 0$. Instead we find
\begin{equation}
G_{\pi\pi}^{-1}(0) = {8y \over v} \int^{\pi\over a}_{-{\pi\over a}}
{d^4k \over (2\pi)^4}
{ a r \khat^2 \over \ktilde^2 + (yv + a r \khat^2)^2 }~,
\label{eq:ward}
\end{equation}
where ${\ktilde^2} ={1\over a^2} \sum_\mu sin^2 (ak_\mu) $ and
$\khat^2 = {1\over a^2} \sum_\mu 4 sin^2(k_\mu a/2)$.
As expected, the violation of the chiral symmetry is proportional to the
coefficient of the Wilson term, $r$. Furthermore, this chiral symmetry
breaking term is UV divergent $G_{\pi\pi}^{-1}(0) \sim 1/ a^3$.
The only way to restore the chiral symmetry is to introduce lower dimensional
counter terms to absorb the UV divergence
\begin{equation}
{\rm counter \  terms} = A \sigma + B (\sigma^2 - {\vec \pi}^2) +
C \sigma (\sigma^2 + {\vec \pi}^2)  + \lambda_2 \sigma^2 {\vec \pi}^2
+ \lambda_3 ({\vec \pi}^2)^2~.
\end{equation}
Note that we can also add a bare fermion mass term $m_0 \psibar\psi$.
However, this can be absorbed by a shift in $A$.

It is easy to show in the large $N$ limit that by tuning the linear counter
term $A$, the leading $1/a^3$ divergence in Eq. (\ref{eq:ward}) can be
canceled.
Similarly, tuning $B$ will cancel the remaining $1/ a^2$ divergence
and $C$ the linear $1/a$ divergence. The conditions on $A, B$ and $C$
are given by
\bea
A &=& 8 y \int {d^4k\over (2\pi)^4} {ar\khat^2 \over \ktilde^2
+ \left(ar\khat^2\right)^2}~,\nonumber\\
B &=& 32 y^2 \int {d^4k\over (2\pi)^4} {(ar)^2(\khat^2)^2 \over
\left(\ktilde^2 + \left(ar\khat^2\right)^2\right)^2}~,\nonumber\\
C &=& - 24 y^3  \int {d^4k\over (2\pi)^4}  {ar \khat^2 \over \left(\ktilde^2
+ \left(ar\khat^2\right)^2\right)^2} + 96 y^3  \int {d^4k\over (2\pi)^4}
{\left(ar \khat^2\right)^3 \over  \left(\ktilde^2 +
\left(ar\khat^2\right)^2\right)^3}~,
\eea
Finally, $\lambda_2, \lambda_3$ should be tuned to remove
any residual $O(1)$ (or $O(log(a))$) symmetry breaking terms.
After such tuning, the chiral Ward identity is satisfied to $O(a)$ and the
chiral symmetry will be restored in the continuum limit. This works exactly
in the same way as the conventional lattice QCD. The only complication is that
now we have three or more counter terms to tune instead of one. Of course, the
advantage here is that the numerical simulation can be done with the counter
terms fixed exactly at the chirality point.

For XQCD, the value of the counter terms have to be determined numerically.
Besides the usual $G_{\pi\pi}^{-1}(0) = 0$ condition, we need additional Ward
identities in order to completely fix the counter terms. This procedure is
similar to a recent proposal for chiral fermions on the lattice
\cite{Italian}. Only there one needs to restore a local gauge symmetry in
contrast to a global chiral symmetry for XQCD.

\newpage
\noindent{\bf VI. Discussion}

The main goal of this formulation of lattice QCD is to try to avoid the source
of critical slowing down due to small current algebra quark masses. In XQCD we
believe that simulations can be done directly at the physical quark (or pion)
mass avoiding a major source of error due to mass
extrapolations. The price one has to pay is to devise matching conditions that
fix the counter terms and the new lattice parameters: $y_0$ for the mixing of
the the elementary and composite lattice pions and the VEV or mass of the
sigma model. Additional parameters are always the price of any ``improved''
action. Although in principle universality can be evoked to remove completely
their impact on the continuum limit, in practice a large degree of freedom
is allowed to use additional parameters to mimic the
low energy physics properties better on the
lattice with a finite lattice spacing. Naive actions, which
set these parameters to zero are really not free of such arbitrary choices.

What appears unusual about our action for XQCD is the
introduction of new ``elementary'' fields.  Usually one regards the
universality class of cut-off actions to be spanned by polynomials in original
``fundamental'' fields, but not composite or collective fields.  However, as we
pointed out in Section II, polynomials in fermionic fields are sometime
equivalent to additional scalars. In fact this idea of universality between
elementary fermions and bosons is not without precedent.

In the case of the NJL model, the four fermion interaction is
trivially equivalent
to a bosonic Lagrange multiplier fields, which may be extended to dynamical
scalars without changing the continuum field
theory \cite{UCSD,Zinn}.  Another interesting example along these
lines is the suggestion that the 2-d Sine-Gordon theory (or its fermionic
equivalent the massive Thirring model) can be represented as a theory with {\bf
both} elementary scalar and fermion fields \cite{DAMGAARD}. This
example has even been suggested as a toy model for the disappearance (or
irrelevance) of the bag constant as a new mass scale \cite{XBAG},
a suggestion for QCD based on similar phenomenological insight as our own.

Ultimately neither the validity or the usefulness of XQCD for quantum
Chromodynamics can be demonstrated by resorting to models with analogous
properties. We have investigated the phase diagram in the strong coupling
limit ($\beta \to \infty$) and found the result consistent with the
phase diagram given in Figure 1. But we feel that only a careful numerical
study of the phase volume as we approach weak gauge coupling region
can give reliable
evidence to support or negate the universality conjecture.

Our numerical simulations have the goal of studying both the toy systems
beyond the large N expansion and full XQCD itself. As the first step, we have
simulated 2-d lattice with two coupled Ising spins with a single $Z_2$
invariance and checked that they have the same critical exponents and low
lying excitations as the single component Ising model, independent of the
coupling $y_0$ between the two spin systems \cite{KOSTAS}.

For lattice XQCD we are formulating both Wilson and stagger fermions, with
counter terms (as discussed above). Our first goal is to examine in detail the
phase boundary at the chiral transition. For bulk quantities we can use small
lattices so the inclusion of fermion loops is not prohibitive.  Fortunately on
the computational side there is also a similarity between XQCD and the
program suggested recently by Myint and Rebbi \cite{MR} to extract an effective
chiral Lagrangian from conventional lattice QCD.  By an analysis of the low
momentum chiral expansion, we will be able to compare the same operators
computed in standard QCD versus XQCD.  One qualitative property of
universality is that this comparison should not depend too sensitively on the
additional parameters introduced in XQCD, once they are tuned to reasonable
values.  We can check this property as well as begin to see if XQCD is
superior for extracting phenomenologically useful results.

{\bf Acknowlegement:} We gratefully  acknowledge many useful discussions with
S. Chivukula, S. Myint, K. Orginos and C. Rebbi.

\pagebreak

\end{document}